# Using a quantitative assessment of propulsion biomechanics in wheelchair racing to guide the design of personalized gloves: a case study.

Félix Chénier[1,2,*], Gerald Parent[3], Mikaël Leblanc[3], Colombe Bélaise[3], Mathieu Andrieux[3]

[1]Mobility and Adaptive Sports Research Lab, Université du Québec à Montréal, Montreal, Canada.

[2]Centre for Interdisciplinary Research in Rehabilitation of Greater Montreal, Montreal, Canada.

[3]INÉDI Research Centre, Centre collégial de transfert de technologie (CCTT) du Cégep régional de Lanaudière à Terrebonne, Terrebonne, Canada.

*Corresponding author:
Félix Chénier
chenier.felix@uqam.ca
Office SB-4455, Pavillon des sciences biologiques
141 Président-Kennedy, Montréal QC, H2X 1Y4
+1-514-987-3000 #5553

## Abstract

This study with a T-52 class wheelchair racing athlete aimed to combine quantitative biomechanical measurements to the athlete's perception to design and test different prototypes of a new kind of rigid gloves. Three personalized rigid gloves with various, fixed wrist extension angles were prototyped and tested on a treadmill in a biomechanics laboratory. The prototype with 45° wrist extension was the athlete's favourite as it reduced his perception of effort. Biomechanical assessment and user-experience data indicated that his favourite prototype increased wrist stability throughout the propulsion cycle while maintaining a very similar propulsion technique to the athlete's prior soft gloves. Moreover, the inclusion of an innovative attachment system on the new gloves allowed the athlete to put his gloves on by himself, eliminating the need for external assistance and thus significantly increasing his autonomy. This multidisciplinary approach helped to prototype and develop a new rigid personalized gloves concept and is clearly a promising avenue to tailor adaptive sports equipment to an athlete's needs.

# Introduction

The first biomechanical studies on wheelchair racing were done during the late 1980s, with the main objective being to enhance the athlete's performance (Ridgway et al., 1988; Sanderson & Sommer III, 1985). Since then, radical technological advances have led to major modifications to the wheelchair. For instance, wheels were reduced from four to three, a crown compensator to assist the athlete in negotiating transitions from straights to curves was added, and lighter frames and wheels are now used (Cooper & De Luigi, 2014). The technical optimization of the wheelchair, the interface and the accessories is in fact a main objective of most biomechanical studies in Paralympic sports (Morriën et al., 2016).

Among the most recent shifts in wheelchair racing optimization is the athletes' choice regarding the gloves used to push the wheels during propulsion. During the last 10 years, most elite wheelchair racers have replaced traditional soft gloves made of padded leather and Velcro with rigid, thermoformed plastics (Rice, 2016). Optimizing the gloves remains a promising avenue, since at every stroke, a large quantity of kinetic energy initially stored in the upper body must be transferred to the pushrim via the gloves, during a very short amount of time (Vanlandewijck et al., 2001). In standard wheelchair propulsion, a large component of the force applied on the pushrims does not contribute to wheelchair motion (Boninger et al., 1999; Robertson et al., 1996; H. W. Wu et al., 1998), and wheelchair racing thus seems to be mechanically inefficient (Chénier et al., 2021). As such, any energy that is lost in deformation, friction or bouncing during this short contact must be avoided to improve propulsion efficiency and thus performance.

Rice et al. (2015) are the only authors to have measured the impact of both types of gloves on temporal and pushrim kinetic parameters using instrumented racing wheels. They observed differences in some kinetic parameters (e.g., braking moment) and spatiotemporal parameters (e.g., cadence, push angle) at submaximal steady state velocities, in 9 athletes competing in two classes (T53 and T54). Measuring the biomechanical parameters related to different gloves has never been done on athletes with more severe disabilities. It is highly probable that the diversity of disabilities and preferences lead to different optimal gloves designs. For instance, a person with affected wrist control may benefit from additional wrist support.

Personalizing racing accessories such as racing gloves is more accessible than ever, with recent technologies such as motion capture, force measurement devices and rapid prototyping devices (e.g., 3D printers). However, while 3D motion capture has largely been used to enhance the positioning, seating interface, or propulsion technique in standard wheelchair propulsion (Dellabiancia et al., 2013) and in court sports wheelchair propulsion (B. S. Mason et al., 2013), to our knowledge no recent literature has focused on optimizing the propulsion in wheelchair racing using 3D motion capture. For kinetic parameters, although some instrumented wheels have been built (Chénier et al., 2021; Goosey-Tolfrey et al., 2001; Limroongreungrat et al., 2009; Miyazaki et al., 2020; Rice et al., 2015), an instrumented racing wheel has never been used as a tool to drive personalized design in wheelchair racing.

In this single case study, we present how combining quantitative biomechanical measurements with a qualitative user perception questionnaire helped us to create and test different prototypes of a new kind of rigid gloves that are adapted to the specific needs of a single athlete with Charcot-Marie Tooth disease.

## Context of the study

This project was initiated when a wheelchair racing athlete contacted the research team about the possibility of designing new racing gloves that would be more tailored to his needs. The athlete reported that his current soft gloves were slipping during propulsion and induced numbing of his hands. He also mentioned that before propelling, he needed assistance of approximately 10 minutes to tape his hands and forearms to keep his wrist joints as stable as possible.

### Participant

The aforementioned person is a regional parasport athlete, of functional class T-52, male, aged 32, 1.75 m, 57 kg. He has had an IPC (International Paralympic Committee) permanent functional classification since the start of his athletic career. The origin of his disability is a neuropathic disease (Charcot-Marie Tooth, type IIa), which impairs the strength of his distal musculature. He has muscular atrophy in his forearms, hands, thighs, legs and feet, and moderate muscular atrophy in his arms. His muscular strength levels scored 3 on the ASIA scale for the proximal musculature of upper and lower limbs and 0 on fingers and toes.

He gave his written consent to take part in the study and to publish the results and conclusions. The protocol was developed in conformity with the ethical principles of Cégep de Terrebonne.

### Gloves design

The design of the gloves aimed to 1) stabilize the athlete's wrists in an optimal position for propulsion, and 2) allow him to put his gloves on and attach them independently. In a first ideation phase, two strap design prototypes were built to test different strapping solutions (Fig. 1a), and a geometry design prototype was built to test the overall geometry of the gloves (Fig. 1b). These prototypes were tested with the athlete to ensure that he can put the gloves on by himself and that they give enough support to his wrist.

| | | |
|---|---|---|
| **(a) Ideation phase:** Strap design prototype | Design 1 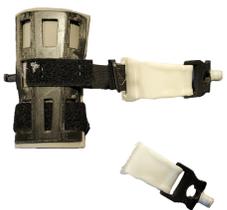 | Design 2 |
| **(b) Ideation phase:** Geometry design prototype | 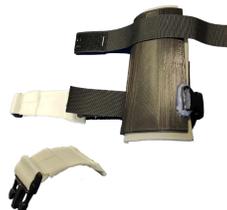 | |
| **(c) Testing phase** | Shown in Fig. 2b | |
| **(d) Follow-up:** Final prototype | 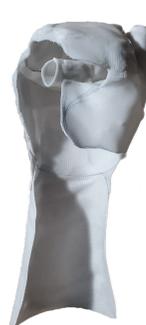 | |

Figure 1. Pictures of the iterative design process

Based on our observations and with the feedback from the athlete, we determined during a second design phase that stability would be increased if the gloves covered the 2/3 of the forearm length and circumference and would be attached by nylon Velcro bands and g-hooks. Moreover, a mushroom shaped button on the back of the hand and rings at the extremity of the straps were included on the gloves to help the athlete to put it on autonomously.

We took 3D scans of the participant's hands and forearms and created 3D-printed prototypes using fused filament fabrication with PLA plastic, as shown in Fig. 2b. To determine which fixed wrist extension angle would be optimal, the participant was asked to wear his usual soft gloves, stabilized with tape, as usually done. We measured his wrist extension in this static, unloaded condition, using a goniometer. This gave an extension angle of 45°, which we selected as the fixed extension for one rigid prototype. However, since the athlete's usual gloves are soft, this 45° angle was expected to vary during propulsion, and therefore we built two other prototypes with lower and higher angles of 30° and 55° extension.

## Prototype evaluation

### Instrumentation

**Gloves:** In addition to the three pairs of prototypes, the participant also used his original racing gloves (soft-cushioned, leather coated and strapped at wrist level) to allow for comparisons.

**Wheelchair and wheels:** The participant used his own custom racing wheelchair (modified Invacare TopEnd). A custom force-sensing instrumented wheel based on the wheel described in Chénier et al. (2021) was installed on the right side of the wheelchair. A 14-inch pushrim, equivalent to the one installed on the participant's left wheel, was installed on the instrumented wheel. The instrumented wheel measured the propulsive moment applied by the athlete on the pushrim at an average frequency of 2.7 kHz.

**Treadmill:** A treadmill (H/P Cosmos, Saturn 300/100r) was used at a 1% incline to simulate the friction effect present in real propulsion conditions (B. Mason et al., 2013).(B. Mason et al., 2013) A guide attached to the treadmill allowed for safe anteroposterior limitation of the wheelchair's movements.

**Motion capture:** The kinematics of the participant and wheel were acquired using a passive 17-camera optoelectronic system (Optitrack, Motive 2.3.0) at a frequency of 180 Hz. Two rigid bodies of 3 reflective markers were affixed to the participant's right forearm and hand as shown in Fig. 2. Five reflective markers were attached to the racing wheelchair's wheel. The medial and lateral epicondyles and the ulna and radial styloid processes were digitized using a probe and expressed in relation to the forearm rigid body. The 2$^{nd}$ and 5$^{th}$ metacarpal heads were also probed and expressed in relation to the hand rigid body.

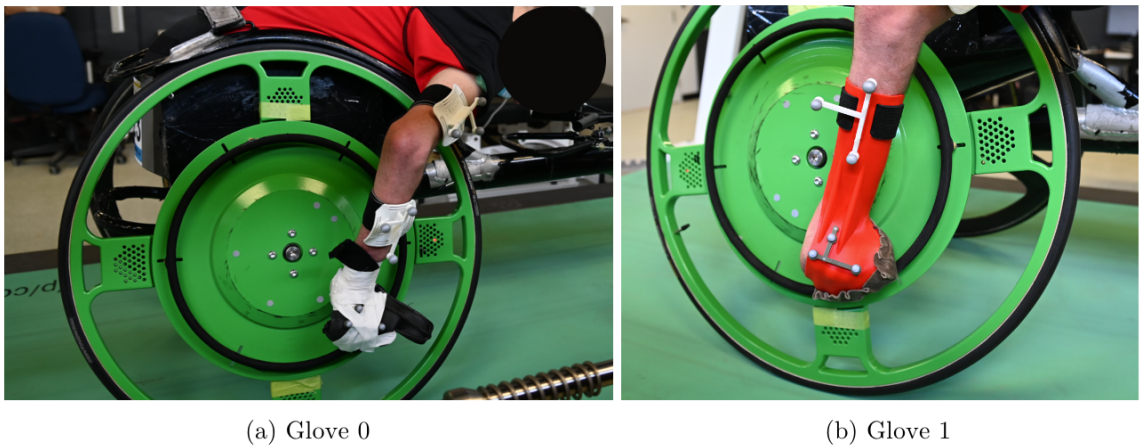

(a) Glove 0          (b) Glove 1

Figure 2. Rigid bodies affixed on the hand and forearm

## Tasks

**Speed selection:** The participant was instructed that he would have to propel himself at a high speed in continuous trials of more than one minute. To determine the propulsion speed that would be used during the tests, he was asked to propel using his own soft gloves while the speed of the treadmill was gradually increased, until he was unable to sustain the treadmill speed (at which point the wheelchair position was moving backward on the treadmill). This speed was 3.5 m/s. The propulsion speed used for all the following tests were set at 2/3 of this max speed, which corresponds to a common training speed, for a speed of 2.3 m/s.

**Gloves testing:** Each pair of gloves was tested twice, including the athlete's own gloves, in the following order: [0, 0, 2, 2, 1, 1, 3, 3], for a total of eight propulsion trials, with gloves 0 being the athlete's own gloves, and gloves 1, 2 and 3 being the 30°, 45° and 55° wrist extension prototypes, respectively. At the beginning of each trial, the participant was stationary on the treadmill. Then, after gently impacting the instrumented wheel to synchronize the wheel to the motion capture system, the treadmill gradually accelerated up to the predetermined speed of 2.3 m/s in less than 10 seconds. After one minute of acquisition at steady speed, the participant was instructed that the acquisition was completed and that he could stop propelling. He was then asked to rate his perceived level of effort using the 6–20 Borg scale for perceived exertion (Borg, 1982), and to rate his level of satisfaction with the gloves on a 0–10 scale. He was also asked to formulate his overall impression of the gloves that included various aspects like comfort, adjustment, and stability. A minimal pause of 10 minutes was allocated between each trial to recover and limit fatigue.

## Data processing

**Kinematic measurements:** The 3D trajectories of the reflective markers in space were filtered at 6 Hz using a no-lag, 2nd order Butterworth low-pass filter. This cut-off frequency was chosen based on the frequency spectrum of wheelchair propulsion kinematics (Cooper et al., 2002).

Wrist extension angle $\theta_{wrist}$ was calculated using the right forearm and right-hand coordinate systems, expressed from the reconstructed bony landmarks in accordance with the recommendations of the International Society of Biomechanics (G. Wu et al., 2005). Wrist extension was defined as the first angle in a sequence of three mobile Euler angles (ZXY).

Hand position angle $\theta_{hand}$ was expressed in a fixed wheel hub coordinate system created from the circular motion of the wheel's markers, with the origin being the wheel centre, x being forward and z being normal to the wheel plane, and y being upward and inward due to wheel camber. $\theta_{hand}$ was defined as the angle between y and a line from the wheel centre to the hand, with 0° being the top of the pushrim and 90° being the pushrim's most forward point (Vanlandewijck et al., 2001).

Wheel rotation angle $\theta_{wheel}$ was calculated using the wheel's markers and was expressed in degrees in the wheel plane.

**Kinetic measurements:** The propulsive moment values measured by the instrumented wheel were filtered at 30 Hz using a no-lag, 2nd order Butterworth low-pass filter. This cut-off frequency was chosen to keep clean transitions during hand-wheel contact and release, to ensure an accurate segmentation of the push phases.

**Cycle segmentation and selection:** Cycles were segmented manually: push phases were defined by the propulsive moment signal being of greater amplitude than the noise floor measured during recovery. The 30 most repeatable cycles were selected based on the similarities of the propulsion moment curves. This selection was done to exclude corrective propulsion cycles due to wheelchair speed adjustment, that happen for instance after the athlete missed a stroke or if the wheelchair is too far on the treadmill. These most repeatable cycles accounted for about 75% of all cycles and were distributed mostly equally in time during the trials.

**Outcome measures:** The following parameters were calculated and averaged over the 30 most repeatable cycles:

- Temporal parameters: push time (s), recovery time (s), cycle time (s)
- Spatial parameters: start angle (deg, defined as $\theta_{\text{hand}}$ at hand contact), end angle (deg, defined as $\theta_{\text{hand}}$ at hand release), push arc (deg, defined as $\theta_{\text{wheel}}(\text{release}) - \theta_{\text{wheel}}(\text{contact})$).
- Kinetic parameters: mean propulsive moment during push phase (Nm), angular impulse (Nm·s, defined as mean propulsive moment × push time).

The entire data processing was performed using Matlab R2022b (Mathworks).

# Results

The perception of each glove by the athlete is shown in Table 1. In terms of user perception, the prototype that gave both the lowest perception of effort and the highest general rating were gloves 2. While the athlete found gloves 2 uncomfortable at first, he got used to it, he felt that his propulsion pattern was healthier for his shoulders, and he liked the increased wrist stability that it provides. He found that the highest wrist stability was attained with gloves 3, but also found that these gloves made it increasingly harder to make good contact with the pushrims. The worsts gloves were gloves 1, as he found that he touched the pushrims too late during the propulsion cycle using these gloves.

Table 1. Perception of the gloves by the athlete

| Gloves | Borg scale (6–20) | General rating (0–10) |
|---|---|---|
| **0 (current soft gloves)** | 12 – light to somewhat hard | 5.0 |
| **1 (30°)** | 15 – hard (heavy) | 4.5 |
| **2 (45°)** | 12 – light/somewhat hard | 5.5 |
| **3 (55°)** | 14 – somewhat hard to hard | 4.5 |

By associating higher wrist stability to lower variation (range) of wrist extension, we can assess wrist stability in Fig. 3 and Fig 4. Based on the measured extension ranges of 18.5° (gloves 0), 5.4° (gloves 1), 5.3° (gloves 2), and 3.3° (gloves 3), we conclude that the three rigid prototypes were effectively stabler than his usual soft gloves, with gloves 3 being the most stable.

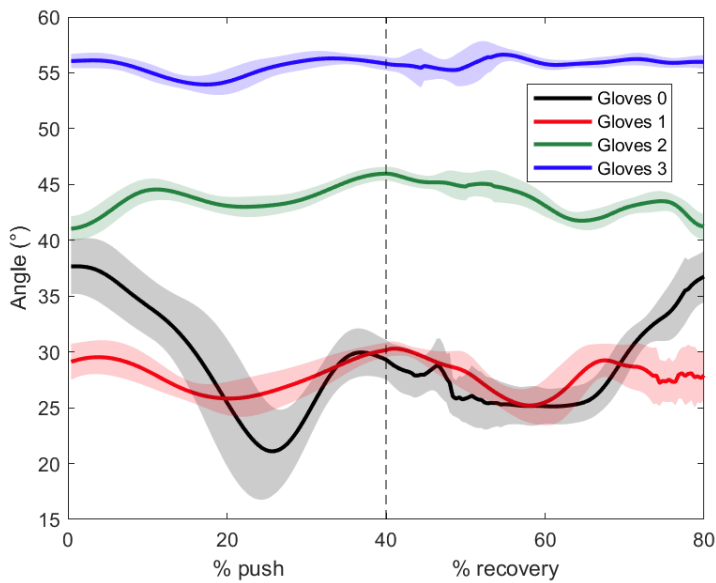

Figure 3. Wrist extension during the push and recovery phases for the four pairs of gloves

We also confirm that for every rigid glove, the wrist extension measured during propulsion (27.8°, 43.7° and 55.4°) was coherent with its initial design (30°, 45°, 55°). However, although the static, unloaded wrist extension with gloves 0 have been measured as 45°, it was much lower in dynamic loaded conditions, with an average value of 29°. This lower extension is most probably due to deformation of the gloves. Therefore, from a kinematic aspect, the prototype with the wrist angle most similar to the athlete's current gloves were gloves 1.

Figure 4 shows the spatiotemporal, kinetic and kinematic parameters for each pair of gloves. The largest differences between the three prototypes were in push time, cycle time, start angle, push arc, impulse and expectedly, average wrist extension angle. Globally, from a temporal and kinetic aspect, the prototype with the most similar parameters to the athlete's current gloves were gloves 2.

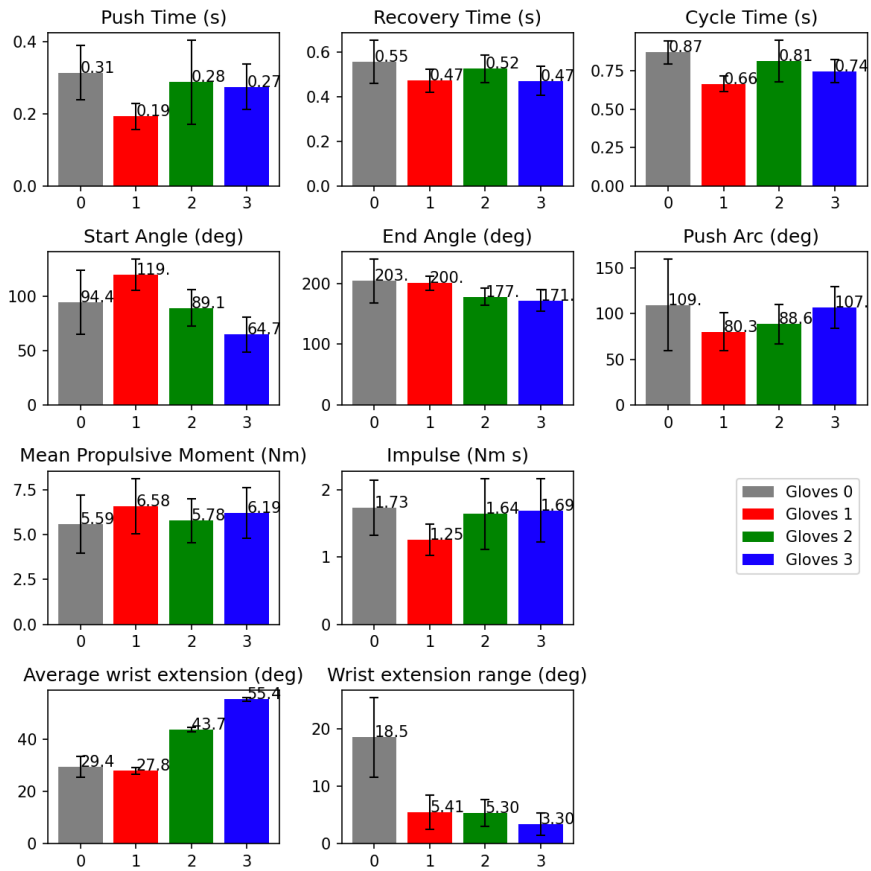

Figure 4. Spatiotemporal, kinetic and kinematic measurements for the four pairs of gloves

Figure 5 shows the evolution of the propulsion moment (a) in time from push initiation, and (b) in hand position angle. Although the mean propulsive moments are similar in Figure 4, we observe in Figure 5a that the propulsive moment reaches a higher peak with gloves 1, and that this peak is reached earlier with gloves 1 and 3 than with gloves 0 and 2. Figure 5a highlights the differences in push time, with gloves 1 being the shortest and gloves 0 being the longest. Figure 5b shows the same moment but as a function of hand position angle. It strongly highlights the spatial differences between the gloves, not only in push arc, but also in start and end angles. While peak moment is reached within about 0.07 to 0.13 seconds after impact for every glove, these instances correspond to very different locations on the pushrim for the different gloves: this peak happens at about 140° for gloves 0 and 2, at 160° for gloves 1, and from 100° to 130° for gloves 3.

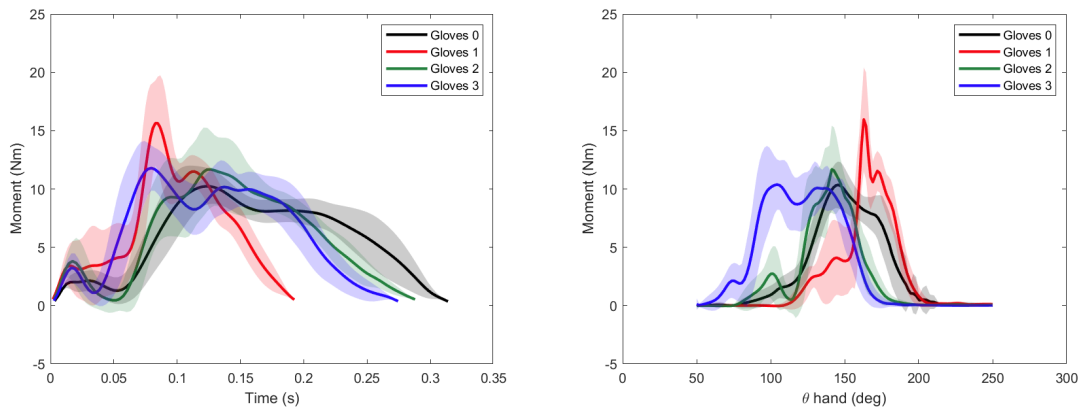

(a) As a function of time  (b) As a function of hand position angle $\theta_{\text{hand}}$

Figure 5. Propulsive moment during the push phase for the four pairs of gloves

Finally, Fig. 6 shows the trajectory of the hand with the four gloves. Overall, the trajectories were similar between gloves, although the trajectory of the hand had a higher amplitude during the recovery phase with gloves 0.

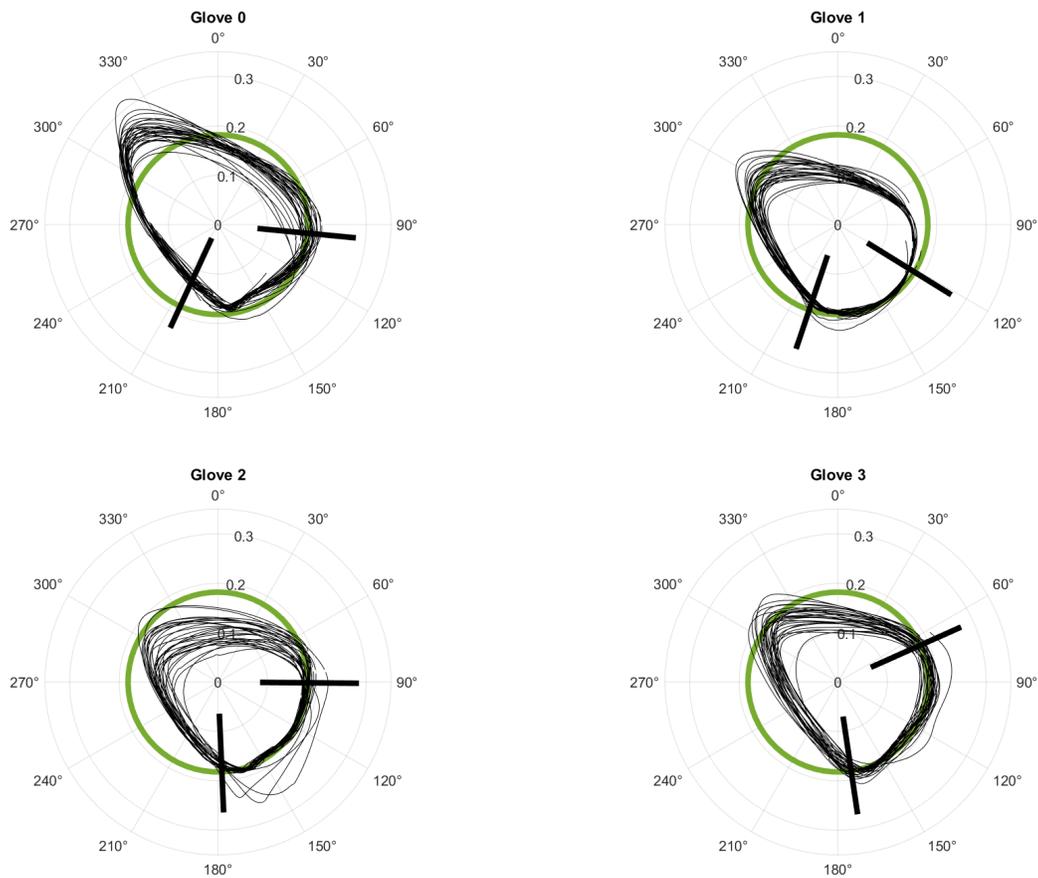

Figure 6. Hand trajectory for the four pairs of gloves

## Discussion

The objective of the study was to investigate how a quantitative assessment of racing wheelchair biomechanics added to a qualitative user perception questionnaire could enhance our understanding of the user's perception and gain new insights into gloves development.

As gloves 1 have a posteriori the most similar wrist extension to gloves 0 for most of the cycle, we would expect that the preference of the athlete would go toward gloves 1. However, the contrary is true, since gloves 1 produced the least satisfaction and the highest perceived effort among the three prototypes. The athlete felt that his hand contacted the wheel further on the pushrim, which is confirmed in Fig. 4 and Fig. 5. In fact, the start angle appears to be strongly related to the wrist extension angle, with the most extended wrist condition making contact sooner with the pushrim.

The athlete's main comment on his favourite prototype (gloves 2) was that he liked the increased stability of the wrist in comparison to his own gloves (gloves 0). Apart from increased stability, which can effectively be observed compared to gloves 0 in Fig. 3 and Fig. 4, we believe the reason for this preference may be explained by his ability, using these gloves, to propel using a very similar technique as with gloves 0. Indeed, most parameters of Fig. 4 and Fig. 5 were the most similar to gloves 0: push time, recovery time, cycle time, start angle, and mean propulsive moment. Not only were these parameters similar, but the peak moment production was at the same time after hand contact, as seen in Fig. 5a. Most notably, the peak moment production was also at the same position on the wheel, as seen in Fig. 5b, which may feel more natural for him compared to the other prototypes that shifted the moment production curve greatly in terms of hand position angle.

Although the athlete preferred gloves 2, he felt that gloves 3 provided the best stability, which is coherent with Fig. 3 and Fig. 4 where the wrist extension varies the least during the propulsion cycle. He also mentioned that he felt that the point of contact was also the most stable. We observe in Fig. 5 that he was able to maintain a high, constant propulsive moment on a longer arc, which may be related to this feeling of hand contact stability. However, while he found this contact easy at the beginning, he felt tired faster, and this may be related to the hand contact that occurred so early during the propulsion phase. Wheelchair racing propulsion technique implies a transfer of kinetic energy from the trunk and arm movements before contact, to the wheel during the push (Vanlandewijck et al., 2001). Contacting the wheel too soon may have decreased the ability of the athlete to generate kinetic energy with the trunk and arms, and therefore increased the muscular demand from the arms.

By closely inspecting Fig. 6, we remark that for gloves 0, the athlete continued pushing the wheel even after releasing the pushrim, from 180° to 200°, which may seem puzzling at a first glance. By carefully inspecting video capture of the athlete, we realized that during this time, the soft gloves were not in contact to the pushrim anymore, but they remained in contact with the wheel via the back of the hand, effectively increasing the push arc. This behaviour was not observed in any of the tested rigid gloves, most probably because the rigid gloves were thinner, harder, and did not provide soft cushioning or grip on the back of the hand. This realization may serve as a possible investigation direction for future work, on the possible use of soft grip on the back of the hand to increase the push arc.

This observation also confirms that the athlete may propel using the wheel surface itself in addition to the pushrim, an assumption that led us to design the instrumented wheel so that the whole surface is instrumented (Chénier et al., 2021). It also highlights the importance of the so-called rotation phase of the racing wheelchair propulsion cycle described by Vanlandewijck et al. (2001), that happens between the push phase and hand release. Finally, it indicates that for this sport, segmenting pushes using cameras can give very different results than using an instrumented wheel. The former indicates when the hand is on the pushrim, while the latter indicates when a propulsion moment is actually applied by the hand.

As a first iteration, this work that combines traditional design and biomechanical assessment already has practical value, because:

1. An initial problem with the athlete's current gloves was solved: he can now put his gloves on by himself, and the gloves slip much less because they are moulded to his hands.
2. It appears that one of the prototypes (gloves 2) allows him to maintain most of his original technique, which is positive since switching to these radically new gloves should not be associated with a loss of performance.

As a follow-up, the athlete mentioned after a few training sessions on the track with his new gloves (gloves 2), that he wished for even better general stability as the gloves were slowly moving on his hands as he propelled on long races. This led to a second iteration with added straps for a better fit on the forearms that he has been using ever since (Fig. 1d**Error! Reference source not found.**).

The data acquired during this first assessment can be used to guide the design of the next iteration of gloves, by reinterpreting it in an aspect of pain prevention. We note in Fig. 5a that among the three prototypes, gloves 1 were the ones with the highest moment rate of rise. We also note that although the mean propulsive moment was similar between gloves, the peak is much higher for gloves 1: as per Fig. 5b, the moment was indeed very low during the 120° to 150°, before spiking at 170°. Finally, this propulsion moment happens in the least amount of time, with a push time of 0.19s vs. 0.31s for gloves 0, and consequently generates the least impulse, with 1.25 Nm·s vs. 1.73 Nm·s for gloves 0. This leads the athlete to increase his cadence, with a cycle time of 0.66 s vs. 0.87 s for gloves 0, to keep up to speed with the treadmill. All these observations go against the recommendations for preservation of upper limb integrity in standard wheelchair propulsion (Consortium for Spinal Cord Medicine, 2005), as they have been correlated to higher risk of developing shoulder and wrist disorders and pain in standard wheelchair propulsion (Boninger et al., 2005; Mercer et al., 2006; Mulroy et al., 2006). Although we should not directly transfer these recommendations from standard wheelchair to racing wheelchair since the technique is so different, it minimally signals that for future iterations of the racing gloves for this athlete, decreasing wrist extension angle should be done with care by taking these possible risks into account.

A similar study by Costa et al. (2009) aimed to personalize the equipment of an athlete using technological instrumentation. The authors aimed to find the best pushrim diameter for one elite athlete of class T52, also diagnosed with Charcot-Marie Tooth disease. They calculated push time and stroke frequency using a high-speed camera, heart rate using a training heart-rate monitor, and lactate using a portable lactate analyzer, for three pushrim diameters. Interestingly, this was, to the authors' knowledge, the only study to describe the use of technological instrumentation as a method to personalize wheelchair racing equipment. As a matter of fact, 3D kinematic instrumentation has been used in labs before (De Klerk et al., 2022; Lewis et al., 2018; Poulet et al., 2022), but mainly to better understand the principles of wheelchair racing propulsion performance and injury prevention. Experimental prototypes of instrumented pushrims have also been developed (Chénier et al., 2021; Goosey-Tolfrey et al., 2001; Limroongreungrat et al., 2009; Miyazaki et al., 2020; Rice et al., 2015); however, this is the first time these instruments were used together to personalize wheelchair racing equipment.

Consequently, this is also the first time the propulsion moments are reported as a function of hand position. We clearly see in Fig. 5b that changing gloves affected both kinematics and kinetics. A similar experimental method could be used to report the effect of other interventions, such as changing pushrim size, wheel camber, athlete positioning, etc. on both kinematics and kinetics. Moreover, we may envision using similar measurements in more complex analyses such as musculoskeletal simulation, with models that simulate the capacities and incapacities of the athlete. Such analyses may be helpful to better understand the relations between hand kinematics and moment production, in the sake of personalizing and optimizing the wheelchair-user interface and the user's propulsion technique.

As a main limitation of our case study, any modification to sports equipment implies adaptation by the athlete, and this type of one-day experiment cannot allow for such adaptation. In the study by Costa et al. (2009), the athlete rotated between the three pushrims during training for three weeks before the biomechanical test to become accustomed to each. However, using different pushrim diameters is less disruptive than testing completely different gloves designs. Moreover, the neuromuscular and physical condition of the athlete may change with time, and the best gloves at a given time may not be the best a year later. The next logical steps are therefore to continue optimizing the gloves with other iterations of this assessment a few months later, most likely using prototypes with finer differences.

Another limitation is the use of stationary instrumentation instead of collecting data directly on a racing track, which has the potential to interfere with the athlete's own propulsion technique. For instance, although the treadmill slope had been adjusted to generate a similar rolling resistance as described in Mason et al. (2013), the main source of resistance in wheelchair racing at high speed is air drag (Hedrick et al., 1990), which means that propelling on the treadmill may have minimized the propulsive moments needed to reach a similar speed on a racing track. Propelling a standard wheelchair increases the stroke frequency for a same speed (Chénier et al., 2018); it is possible that similar behaviour would be observed for wheelchair racing. Additionally, as seen in Fig. 2, the instrumented wheel has a small bump in its centre, to accommodate its force cell. The athlete indicated that he inadvertently touched it with his gloves on some occasions. However, this was sporadic and we do not believe that his propulsion pattern was affected. These limitations were unavoidable to measure the 3D kinematics and kinetics of the athlete, and to avoid external sources of bias such as variable weather conditions. They do not limit the results of the comparisons between the four gloves, because all gloves were tested under similar conditions. It may, however, impact the transfer of those measurements to real conditions, and this is why continuous follow-up with the athlete is necessary as he trains on a track with his new gloves.

Finally, while this methodology allowed comparing the biomechanical parameters associated to each prototype to the athlete's usual soft gloves, it could hardly be used to directly assess glove performance. Indeed, performance is strongly related to the power developed by the athlete; however, as both the speed and resistance were controlled and constant, then the mean power is also controlled. Controlling these variables was required to observe the biomechanical impact of the gloves only, excluding the possible impact of an associated speed gain. To measure the total performance associated to given gloves, a similar experiment should be done in simulated or real race conditions at maximal speed.

## Conclusion

In this paper, we presented a method to personalize the design of wheelchair racing equipment, namely the conception of new gloves, that adds quantitative biomechanical assessment to traditional iterative design based on qualitative interactions with the user. Such user-centred, personalized design is important in adaptive sports because the abilities and inabilities between different athletes are so unique. In this case study, we created three variants of rigid gloves that would allow the athlete to be autonomous and to overcome wrist mobilizer weakness due to his disease. The combination of kinematic and kinetic instrumentation allowed to better understand why the user preferred a particular pair of gloves prototype, and will be helpful for designing other iterations as the athlete's physical condition and technique change over time.

## Acknowledgment

This work was supported by the Natural Sciences and Engineering Research Council of Canada.

## Disclosure of interest

The authors report there are no competing interests to declare.